\documentclass[10pt]{article}
\usepackage{amssymb}

\newcommand{\ncm}{\newcommand}
\ncm{\beq}{\begin{equation}}    \ncm{\enq}{\end{equation}}
\ncm{\bea}{\begin{eqnarray}}    \ncm{\eea}{\end{eqnarray}}
\ncm{\barr}{\begin{array}}      \ncm{\earr}{\end{array}}
\ncm{\tfr}{\textstyle\frac}     \ncm{\ed}{\tfr{1}{3}}
\ncm{\mod}{\ \ {\rm mod}\ }     \ncm{\tx}{\textstyle}
\ncm{\ny}{\nonumber}            \ncm{\tf}{\tx\frac} 
\ncm{\Lb}{\left[}               \ncm{\Rb}{\right]}       
\ncm{\lb}{\left\{}              \ncm{\rb}{\right\}}      
\ncm{\lk}{\left(}               \ncm{\rk}{\right)}       
\ncm{\cA}{{\cal A}}      \ncm{\cF}{{\cal F}}      \ncm{\cG}{{\cal G}}           
\ncm{\cH}{{\cal H}}      \ncm{\cJ}{{\cal J}}      \ncm{\cQ}{{\cal Q}}
\ncm{\cZ}{{\cal Z}}      \ncm{\Zmb}{\mathbb{Z}}   \ncm{\Cmb}{\mathbb{C}} 
\ncm{\alp}{\alpha}       \ncm{\om}{\omega}        \ncm{\sig}{\sigma}   
\ncm{\hs}{\hspace*{1cm}} \ncm{\hq}{\hspace*{6mm}} \ncm{\hx}{\hspace*{3mm}} 
\ncm{\PQL}{P_Q^{(L)}}    \ncm{\Pic}{\Pi_Q^{(L)}}  \ncm{\hal}{{\tfr{1}{2}}}     
\begin{document}
\begin{center}
{\LARGE\bf Onsager's algebra and partially\\[3mm] orthogonal polynomials}
\\[6mm]{\large\bf G. von Gehlen}\\[2mm]
Physikalisches Institut der Universit\"at Bonn\\
  Nussallee 12, 53115 Bonn, Germany\\
  {\it e-mail: gehlen@th.physik.uni-bonn.de}\\[5mm]\end{center}\vspace*{2mm}
\noindent{\large\bf Abstract:}\\[1mm]
The energy eigenvalues of the superintegrable chiral Potts model are
determined by the zeros of special polynomials which define finite 
representations of Onsager's algebra. The polynomials determining the 
low-sector eigenvalues have been given by Baxter in 1988. In the $\Zmb_3-$case 
they satisfy 4-term recursion relations and so cannot form orthogonal 
sequences. However, we show that they are closely related to Jacobi 
polynomials and satisfy a special "partial orthogonality" with respect 
to a Jacobi weight function.
\\[3mm] {\it PACS:} 02.30.I, 75.10.J, 68.35.R
\\{\it Keywords:} Chiral Potts model, Integrable quantum chains, 
Onsager's algebra     

\section{Introduction} 
F.Y.Wu and Y.K.Wang \cite{FYWu} were the first to consider the Potts model
with chiral interaction terms. Their interest in this generalization arose
from duality considerations, but the idea proved to be very fruitful in
many respects: 
Ostlund \cite{Ost} and Huse \cite{Huse} proposed the chiral Potts (CP) model 
for phenomenological applications: it allows to describe incommensurate 
phases using nearest neighbor interactions only. 
We give a few references \cite{HuFi,Cen,HKDN,Face} from 
which the subsequent development can be traced, and turn directly to the 
superintegrable chiral $\Zmb_N$ Potts quantum chain \cite{GeRi}. 
This is a particularly interesting model, because it provides some of the 
rare representations known for Onsager's algebra \cite{Onsa} and in this sense 
generalizes the Ising quantum chain (for $N=2$ it is the Ising 
model). Integrability by Onsager's algebra entails that all eigenvalues 
of the hamiltonian are determined by the zeros of certain polynomials, 
which for the chiral Potts model were first derived by Baxter \cite{Ba88}. 
Although the definition of Baxter's polynomials looks
very simple, the properties of these polynomials turn out to be quite 
non-trivial and interesting \cite{GeRo}. The main part of this note deals with
the properties of these polynomials. They satisfy $N+1$-term recursion 
relations, therefore for $N>2$ they cannot form orthogonal sequences. However,
as found recently \cite{GeRo}, several properties which characterize 
orthogonal polynomials are almost true for Baxter's polynomials (e.g. the zero 
separation property is true except for one extreme zero).  
\\[2mm]          
We first recall the definitions of the superintegrable CP-hamiltonian and 
Onsager's algebra and then, following B.Davies \cite{Da90}, we sketch how the 
formula for the energy eigenvalues emerges. We consider Baxter's polynomials 
and their recursion relations. Equivalent polynomials with their zeros in
$(-1,+1)$ for $N=3$ are written in terms of a determinant. Their expansion in 
terms of Jacobi polynomials gives the surprising result that many of the 
expansion coefficients vanish, leading to the notion of "partial 
orthogonality". 
\\[2mm] 
The hamiltonian defining the $\Zmb_N$-superintegrable chiral Potts 
quantum chain \cite{HKDN,GeRi} is:
\beq\cH\;=\;-\,\sum_{j=1}^L\sum_{l=1}^{N-1}\frac{2}{1-\omega^{-l}}\lk X_j^l
\;+\: k\:Z_j^l Z_{j+1}^{N-l}\rk. \label{SCP}\end{equation} Here  
$\omega=e^{2\pi\,i/N}$ and $Z_j$ and $X_j$ are $\;\Zmb_N$-spin operators acting 
in the vector spaces ${\mathbb{C}}^N$ at the sites $j=1,2,\ldots,L\;$ ($L$ is 
the chain length). The operators obey $\;Z_i\,X_j\,=\,X_j\,Z_i\,
\omega^{\delta_{i,j}};\hx Z_j^N= X_j^N=1$ and we assume $X_{L+1}=X_1$ 
(periodic b.c).
A convenient representation is $(X_j)_{l,m}=\delta_{l,m+1}\mod N$ and
$(Z_j)_{l,m}=\delta_{l,m}\omega^m.$
For $N\ge 3$ the complex coefficients make the chain hamiltonian parity 
non-invariant. For $N=2$ we get the Ising quantum chain. 
For fixed $N$ there is only one parameter, the temperature variable $k$. 
Incommensurate phases arise due to ground state level crossings. $\cH$ commutes
with the $\Zmb_N$-charge $\;\widehat{Q}=\prod_{j=1}^L\;Z_j$. We write the 
eigenvalues of $\;\widehat{Q}\;$ as $\omega^Q$. $Q=0,1,\ldots,N-1$ labels the 
charge sectors of $\cH$.
\\[2mm]
We split $\cH$ into two operators writing $\; \cH\;=-\hal\,N(A_0+k A_1).\:$
A remarkable property of $\cH$ is that $A_0$ and $A_1$ satisfy \cite{GeRi} 
the Dolan-Grady \cite{DoGr} relations
$$  \Lb A_0,\Lb A_0,\Lb A_0,A_1\Rb\,\Rb\,\Rb\,=\,16\Lb A_0,A_1\Rb;
\hs\;\Lb A_1,\Lb A_1,\Lb A_1,A_0\Rb\,\Rb\,\Rb\,=\,16\Lb A_1,A_0\Rb, $$
which are the conditions\cite{Pe87} for $A_0$ and $A_1$ to generate Onsager's 
algebra $\cA$, which is formed \cite{Onsa} from elements 
$A_m,\,G_l,\;m\in \Zmb,\;l\in {\mathbb N},\;\:l\ge m,$ 
satisfying \beq   [A_l,A_m] = 4\,G_{l-m};\hq
[G_l,A_m]= 2\,A_{m+l}-2\,A_{m-l};\hq  [G_l,G_m]= 0. 
\label{Ons}\end{equation}
From (\ref{Ons}), there is a set of commuting operators which includes $\cH$:  
$$  Q_m=\hal\lk A_m+A_{-m}+k(A_{m+1}+A_{-m+1})\rk;\hx\hx 
     \Lb\: Q_l,\:Q_m\Rb=0;\hx\hx Q_0=\cH. $$   
To obtain finite dimensional representations of $\cA$ we require the $A_m$   
(and analogously the $G_l$) \cite{Da90,DaRo}$\;$ to satisfy a finite 
difference equation:  $\;\;\sum_{k=-n}^n\alpha_k\,A_{k-l}\;=0.$
This is solved introducing the polynomial (the main object of the 
present paper):
\beq   \cF(z)=\sum_{k=-n}^n\alpha_k\:z^{k+n}     \label{zpoly}\end{equation} 
(from $\cA$ the $\alpha_k$ are either even or odd in $k$).
Now the $A_m$ and $G_m$ can be expressed in terms of the zeros $z_j$ of 
$\cF(z)$ and the set of operators $\:E_j^\pm,\;H_j$: \beq 
A_m=2\sum_{j=1}^n\:\lk z_j^m\,E_j^+ +z_j^{-m}
 E_j^-\rk;\hs G_m\:=\:\sum_{j=1}^n\:\lk z_j^m-z_j^{-m} \rk \:H_j.\label{cou}
  \end{equation}
From $\cA$ these operators obey $sl(2,C)$-commutation rules:
$$ [E_j^+,E_k^-]\:=\:\delta_{jk}\:H_k; \hs\hx
    [H_j,\:E_k^\pm]\:=\:\pm 2\,\delta_{jk}\:E_k^\pm.   $$ 
So $\cA$ is isomorphic to a subalgebra of the loop algebra of
a sum of $sl(2,C)$ algebras.
\\[1mm]
From the first of eqs.(\ref{cou}) we can express $\cH$ in terms of the $z_j$
and the operators $E_j^\pm$.
Writing $\;E_j^\pm=J_{x,j}\,\pm\,iJ_{y,j}$, then in a representation $\cZ(n,s)$ 
characterized by the polynomial zeros $z_1,\ldots,z_n$ and  
a spin-$s$ representation $\vec{J}_j^{(s)}$ of all the $\vec{J}_j$, we get:
\bea (A_0+k A_1)_{\cZ(n,s)}&=&2\,\sum_{j=1}^n\lb (2+k\,(z_j+z_j^{-1}))
J_{x,j}^{(s)}+i(z_j-z_j^{-1})\,J_{y,j}^{(s)}\rb \ny\\ 
  &=&4\,\sum_{j=1}^n\,\sqrt{1+2k\,c_j+k^2}\;{J'}_{x,j}^{(s)}\ny\eea
where ${J'}_{x,j}$ is a rotated $SU(2)$-operator, and
$\;\; c_j\:\equiv\;\cos{\theta_j}\;=\:\hal(z_j+z_j^{-1}).$\\[2mm]
For the CP-hamiltonians (\ref{SCP}) the spin representation turns out to 
be $s=\hal.$ Accordingly, all eigenvalues of (\ref{SCP}) have the form 
\beq E\;=\,-N\,\lk\,a\,+\,b\,k\,+\,2\,\tx\sum_{j=1}^n m_j\,
 \sqrt{1+2\,k\cos{\theta_j}\,+\,k^2}\,\rk,\hx m_j=\pm\hal.  
\label{EE} \end{equation} 
$a$ and $b$ are non-zero if the trace of $\:A_0\:$ and $\:A_1\:$ is non-zero. 

\section{Baxter's polynomials}
No direct way is known to find the polynomials $\cF(z)$ from the hamiltonian
$\cH$. However, the invention of the two-dimensional integrable CP 
model \cite{AuYa,BaAu}, which contains $\cH$ as a special logarithmic 
derivative, and functional relations for its transfer matrix have enabled 
Baxter \cite{Ba88} to obtain the polynomials for the simplest 
sector of $\cH$, which at high-temperatures contains the ground state 
(the polynomials corresponding
to all other sectors have been obtained subsequently in \cite{Al89,Dasm,Baxt}). 
Here we shall consider only the simplest case. Baxter \cite{Ba88} finds that
in terms of the variable $\:t\:$ or $\;s\equiv 
t^N=(c-1)/(c+1)\;$ $\;$ (recall $\:c=\cos{\theta}\:$ of (\ref{EE}))$\:$ 
these polynomials take 
the form      \beq \PQL(s)=\frac{1}{N}\sum_{j=0}^{N-1}
         \lk\frac{1-\,t^N}{1-\!\om^j\, t}\rk^L(\om^j\,t)^{-\sig_{Q,L}};
\hx\; \sig_{Q,L}=(N\!-1)(L+Q)\! \mod N. \label{Bscc}\end{equation}   
Here $\;Q\;$ denotes the $\Zmb_N$-charge sector.$\;$ 
For $\;\Zmb_3\;$ eq.(\ref{Bscc}) is, written more explicitly:
$$ P_Q^{(L)}(s)=\frac{t^{-\sig_{Q,L}}}{3}\lb (t^2+t+1)^L
   +\om^Q(t^2+\om^2 t+\om)^L+\om^{-Q}(t^2+\om t+\om^2)^L\rb. $$
Due to their $\Zmb_N$-invariance $\; t\rightarrow \om\,t,\;$ the $P_Q^{(L)}$ 
depend only on $\;s=t^N$. The degree of the $\PQL(s)$ in the variable $s$ is 
$\; b_{L,Q}=\left[((N-1)L-Q)/N \right]\;$
where $[x]$ denotes the integer part of $x$. 
Considering sequences of these polynomials for fixed $Q$ and 
$L \in \mathbb{N}$, we notice that the dimensions $b_{L,Q}$ do not 
always increase by one when increasing $L$ by one: at every $Nth$ step the 
dimension stays the same: e.g. the dimensions of the $P_{Q=0}^{L}$ for 
$L\!\! \mod N=0$ and $L\!\! \mod N=1$ coincide, see Table 1.\\[2mm]
The polynomials (\ref{Bscc}) have their zeros all on the negative real 
$s$-axis: $\cH$ is hermitian and so in (\ref{EE}) we must have 
$\;-1 \le c_j \le+1 \;$ which means negative $s_j$. We will prefer to
deal mostly with equivalent polynomials in the variable $c$, defining
\beq \Pi_Q^{(L)}(c)=(c+1)^{b_{L,Q}} P_Q^{(L)}(\, s=\,\tfr{c-1}{c+1}\:).
\end{equation}
Our main concern in this paper is to learn about the properties of the 
$\Pi_Q^{(L)}(c)$ or $P_Q^{(L)}(s)$, e.g. whether these can be arranged into
orthogonal sequences etc. We will find that the $\Pi_Q^{(L)}(c)$ are 
polynomials with quite remarkable properties. A number of special 
features of the $\Pi_Q^{(L)}$ have been discussed recently \cite{GeRo}. 
Here we give some more detailed results for the 
$\Zmb_3$-case. As the recursion relations for the $\Pi_Q^{(L)}$ contain 
a lot of information, we now show how to obtain these.

\section{Recursion relations}
We start with the observation that the coefficients of the $\PQL(s)$ can be 
obtained from the expansion of $(1+t+t^2+\ldots+t^{N-1})^L$, simply 
by taking every $N-$th term of the expansion, starting with the coefficient 
of $\;t^{(N-1)L-Q}.\;$ More precisely, we claim that we can
define the $\PQL$ by the decomposition 
\beq (1+t+t^2+\ldots+t^{N-1})^L=\!
   \lk\frac{\,1-t^N}{1-t}\rk^L=\sum_{Q=0}^{N-1}t^{\sig_{L,Q}}\!\PQL(s)
  \label{Pnth}\end{equation}
demanding the $\PQL$ to depend on $s=t^N$ only.$\;$ Proof: $\;$ Insert 
(\ref{Bscc}) into (\ref{Pnth}) to get:  $$ \lk\frac{\;1-t^N}{1-t}\rk^L=
(1-t^N)^L\;\frac{1}{N}\;\sum_{Q=0}^{N-1}\sum_{j=0}^{N-1}\;
\frac{\om^{j(L+Q)}}{(1-\om^j\:t)^L}. $$
Interchanging the $Q$- and $j$-summations we see that the $Q$-summation
gives zero for $j\neq 0$, leaving only the $j=0$ term, 
which is (\ref{Pnth}).\\
Eq. (\ref{Pnth}) can now be used to obtain recursion relations: Write
\beq (1+t+t^2+\ldots+t^{N-1})\sum_{Q=0}^{N-1}\;
t^{\sig_{L,Q}}\:\PQL(s)\;=\;
    \sum_{Q=0}^{N-1}\;t^{\sig_{L+1,Q}}\:P_Q^{(L+1)}(s). \end{equation}     
Comparing powers of $t,\;\;$ e.g. for $\;L\mod N=0\;$ this gives
\beq \lk\barr{c} P_0^{(L+1)} \\ P_1^{(L+1)}\\ P_2^{(L+1)}\\ \vdots \\ 
   P_{N-1}^{(L+1)}\earr\rk=
  \lk\barr{ccccc}\;\; 1\;\;&\;\; 1\;\;&\;\;\; 1\;\;&\; \ldots \;&\;\; 1\;\;\\ 
   1& s& 1&\ldots& 1
   \\ 1& s& s&\ldots& 1\\ \vdots & \vdots & &\vdots \\ 1& s& s&\ldots & s
 \earr \rk\lk\barr{c} P_0^{(L)}\\ P_1^{(L)}\\ P_2^{(L)}\\ \vdots 
 \\ P_{N-1}^{(L)} \earr \rk, \label{k3} \end{equation}
For $L\mod N=k$ replace $P_Q \rightarrow P_{Q-k}$ cyclically in both 
column vectors, keeping the same square matrix. Recursion relations 
not coupling $P_Q^{(L)}$ with different $Q$ follow by the $N-$fold application 
of these relations, leading to $N+1$-term recursion formulae. These can be 
transcribed into the corresponding formulae for the $\Pic.$ For the Ising case 
$\Zmb_2$ these are of the Chebyshev type
\beq\Pi_Q^{(L+4)}-4c \Pi_Q^{(L+2)}+4\Pi_Q^{(L)}=0,\label{cheb}\end{equation}
and so for $N=2$ the $\Pi_Q^{(L)}$ form orthogonal sequences. However,
for $\Zmb_3$ we have \cite{GeRo} (valid for all $L\ge 0$ and all $Q$): \beq  
\Pi_Q^{(L+9)}-3(9c^2-5)\Pi_Q^{(L+6)}+48\,\Pi_Q^{(L+3)}-64\,\Pi_Q^{(L)}=0.
\label{rik}\end{equation} 
These $\Pic$ form 9 sequences, each labeled by $(L_0,Q)$, where $Q=0,1,2$ and 
$L=3j+L_0$ where $L_0=0,1,2,\;\;$ $j=0,1,2,\ldots$.
The degrees of the polynomials appearing in this relation increase by two from
the right to the left, but since the recursion is four-term, not three-term, 
these are not orthogonal sequences \cite{Chih}\footnote{If we consider 
$Q=L\!\mod 3$, then we have only polynomials in $c^2$, and the degrees 
$b_{L,Q}$ are consecutive in powers of $z=c^2$ ("simple sets of polynomials") 
with integer coefficients.}. 
However, like (\ref{cheb}) also (\ref{rik}) are of the most simple 
type\footnote{For higher $N$ we get 
similar $N+1$-term relations, e.g. for $\Zmb_4$:\\ 
$\Pi_Q^{(L+16)}-4(64c^3-56c)\Pi_0^{(4)}\Pi_Q^{(L+12)}-128
 (14c^2-17)\Pi_Q^{(L+8)}-2048c\Pi_Q^{(L+4)}+4096\,\Pi_Q^{(L)}=0.$}: 
all coefficients are independent of $L$. 
\section{Expansion in terms of Jacobi polynomials}
As the zeros of our $\Pic$ are confined to and dense in the interval $(-1,+1)$ 
we call this the basic interval like for orthogonal polynomials. 
Trying to determine (numerically) a weight function by the ansatz 
$\;\int_{-1}^1(1+c)^\alpha(1-c)^\beta c^k\Pic(c)=0\;$ for $\;k<b_{Q,L}\;$ 
fitting $\alpha$ and $\beta$, we find (in the following we concentrate on the 
$\Zmb_3$-case) that there is an approximate solution very close 
to $\alpha=-\beta=\tfr{1}{3}$, but $\alpha$ and $\beta$ come out to be 
slightly $L$ and $k$-dependent, in contrast to what is needed for 
orthogonality. However, for $L \rightarrow\infty$ and small fixed $k$, 
the solutions converge 
towards $\alpha=-\beta=\tfr{1}{3}.\;$ So the $\Pic$ seem to be close to Jacobi 
polynomials $P_k^{(\frac{1}{3},-\frac{1}{3})}$, but can we formulate
an exact relation valid for finite $L$? 
Is there an exact property of the $\Pic$ which replaces orthogonality?
\\[2mm] 
Numerical calculations \cite{GeRo} gave the surprising result 
that seemingly complicated $\Pic$ can be written as a combination of just very 
few Jacobi polynomials, e.g. \\[1mm]$\;\;\Pi_1^{(12)}\,=\, 
3^{11}c^7+3^{10}c^6-5\cdot 3^{10}c^5-80919c^4
  +140697c^3  +27459c^2-16839c-1365\,=
\tfr{6^8}{728}\lk \tfr{63}{22}\:P_7^{(\frac{1}{3},-\frac{1}{3})}
  -\,P_5^{(\frac{1}{3},-\frac{1}{3})} \rk.\;$
For polynomials $\pi(c)$ of degree $n$ we use\footnote{For Jacobi series 
as a generalization of Taylor series, 
see Ch.7 of Carlson \cite{Carl}. We use the standard normalization
of Jacobi polynomials, see e.g. Rainville \cite{Rain}.}: 
\beq \pi(c)\:=\;\Lb\, \pi_0,\:\pi_1,\:\ldots,\:\pi_n\,\Rb\;
         \equiv\:\sum_{k=0}^n \pi_k 
   \;P_k^{(\frac{1}{3},-\frac{1}{3})}(c)    \label{comp} \end{equation}
and define a scalar product with Baxter's variable  
$\;t=((1-c)/(1+c))^{1/3}\;$ (see (\ref{Bscc})) as the weight function (here 
we will not need to specify the normalization):
$$\langle\pi^{(1)}|\pi^{(2)}\rangle=\!\int_{-1}^{+1}\!\!\! dc
 \lk\frac{1-c}{1+c}\rk^{\!\!1/3}\!\!\!\pi^{(1)}(c)\pi^{(2)}(c)=
\!-\!\int_{-\infty}^0\!\!dt\frac{6s}{(1-s)^2}\pi^{(1)}(c(s))\:\pi^{(2)}(c(s)).
  $$
The second part of this definition shows that it makes sense also if 
we prefer to use polynomials in the variable $s$, and that it 
preserves the original $\;\Zmb_3-$symmetry.\\[1mm]
\renewcommand{\arraystretch}{1.3}
\begin{table}[t]\begin{center}
\caption{Examples of $\Zmb_3-$polynomials $\Pi_Q^{(L)}(c)$
and their Jacobi-components defined in (\ref{comp}).}\vspace*{1mm}
\begin{tabular}{rll} \hline
\multicolumn{3}{c}{$\Zmb_3,\hs\hs\hs  Q=L+1 \mod 3$}\\ \hline
$L$  &  $\hs\hs\hs\Pic(c)$  &  $\hq 2^{-[2L/3]}\,\Pic(c)$\\ \hline
3&$9\,c+3        $&$[0,\frac{9}{4}]$\\
4&$27\,c^2-18\,c-5$&$[3,-{\frac {27}{4}},\frac{9}{2}]$\\
5&$81\,\,c^3+27\,c^2-57\,c-11$&
 $[0,-\frac{3}{2},0,{\frac {81}{20}}]$\\
6&$3^5\,c^3+81\,c^2-135\,c-21$&$
[0,0,0,{\frac {243}{40}}]$\\
7&$3^6\,c^4-2\cdot 3^5\,c^3-540\,c^2+270\,c+43$&$
[3,-{\frac {27}{4}},{\frac {171}{14}},-{\frac {729}{40}},
                           {\frac {729}{70}}]$\\
8&$3^7\,c^5+3^6\,c^4-2754\,c^3-702\,c^2+711\,c+85$&
                       $[0,0,0,-{\frac {27}{5}},0,{\frac {243}{28}}]$\\
9&$3^8\,c^5+3^7\,c^4-7290\,c^3-1782\,c^2+1593\,c+171$&
                       $[0,0,0,-{\frac {81}{40}},0,{\frac {729}{56}}]$\\
10&$3^9\,c^6-2\cdot 3^8\,c^5-25515\,c^4+14580\,c^3$&
              $[3,-{\frac {27}{4}},{\frac {135}{14}},-{\frac {243}{20}},
{\frac {9477}{385}},$\\ &$\hs\hs\hs+7965\,c^2-3186\,c-341$&
                       $\hs\hs-{\frac {3^7}{56}},{\frac {3^8}{308}}]$
\\  \hline  \end{tabular}\end{center}\end{table}
\noindent Since the $\Pic$ satisfy the recursion relations (\ref{rik}) 
they can be written as determinants of band matrices with a bottom line 
specifying the initial conditions (which are the lowest $L$ polynomials). 
Omitting the bottom line, we define $j\times j-$band matrices and their 
determinants 
$\;\;R_j=\:det\,\left|\,band\,(\,[\,64,\,48,\,3p,\,1,\,0\,],\,j)\right|\;$, 
e.g. \renewcommand{\arraystretch}{1.4}  
$$ R_6\;=\;\left|\begin{array}{cccccc}\;3p\;& 1& 0& 0& 0& 0\\ 
 48&\;3p\;&1&0&0&0\\ 64 & 48 &\;3p\;&1&0&0\\  0&64&48&\;3p\;&1&0
 \\  0& 0&64&48&\;3p\;&1\\  0&0&0&64&48&\;3p\; \end{array}\right|. $$
\renewcommand{\arraystretch}{1.0} 
where $p=9c^2-5,\,$ so that the polynomials $R_j$ depend only 
on $c^2.$ Now we get the nine sequences of the $\Pic$  
as linear combinations of the $R_j$ which satisfy appropriate initial 
conditions. Abbreviating $\;Q_j=R_j+8R_{j-1}\;$ these are found to be:\\
\beq \begin{array}{ll} 
\Pi_0^{(3j)}=\tf{1}{3}\lk Q_j-8Q_{j-1}+16Q_{j-2}\rk;&
\Pi_{1\atop 2}^{(3j)}=9cR_{j-1}\pm 3Q_{j-1};\\[2mm]
\Pi_{0\atop 2}^{(3j+1)}=\pm 18cR_{j-1}+Q_j+2Q_{j-1};&
\Pi_{1}^{(3j+1)}=Q_j-4Q_{j-1};\\[2mm]
\Pi_{0\atop 1}^{(3j+2)}=3c(R_j-4R_{j-1})\pm(Q_j-4Q_{j-1});\;\;\hx&
\Pi_2^{(3j+2)}=3Q_j.\end{array}\label{kvs}\end{equation}  
\vspace{1mm}  For $j=1,2$ use $R_0=1;\;\:R_{-1}=R_{-2}=0.$ 
From (\ref{kvs}) we see that only two of the 9 sequences are independent.
There are relations like e.g. 
$\Pi_0^{(3j+1)}\!\!-\!\Pi_1^{(3j+1)}=2\Pi_1^{(3j)}$. \\[2mm]
To get the Jacobi-expansion of the $\Pi_Q^{(L)}$, we only need to expand 
$R_j$ and $cR_j$. Using Jacobi-components defined in (\ref{comp}),
from explicit calculation (for $j\le 36$), we find that for $\:k<j \:$ 
({\it only there}) 
the $j$-dependence of the ${(R_j)}_k\;$ obeys the simple rule:
$${(R_j)}_k\;=(-3)^k\,k!\;(2k+1)\lb (-8)^j\:\sig_k\;+\:4^j\:\tau_k\rb;$$
$$ \tau_k=\frac{1}{3\prod_{n=0}^k\:(3n+1)};\hx\;\;
\sig_{k=2m}=\frac{2\,(-3)^m}{3\prod_{n=m}^{3m}(2n+1)};\;\;
\hx \sig_{k=2m-1}=\frac{2\,(-3)^{m-2}}{\prod_{n=m}^{3m-1}(2n)}.$$
It follows that the $\sig_k$ do not contribute to the 
$k<j$-components of $Q_j$, and, using the recursion relations for the 
$P_k^{(\frac{1}{3},-\frac{1}{3})}(c)$, we conclude that for $k\le j\;$ 
we have $$\tfr{1}{3}{(Q_j)}_k=-{(cR_j)}_k=\tfr{1}{4}{(c^2 R_{j+1})}_k=
4^j (-3)^k\,k!\;(2k+1) \tau_k.\;\;$$ 
It further follows that  
$${(\Pi_0^{(3j+3)})}_k={(\Pi_1^{(3j)})}_k={(\Pi_0^{(3j+1)})}_k=
{(\Pi_{0\atop 1}^{(3j+1)})}_k={(\Pi_{0\atop 1}^{(3j+2)})}_k=0
 \hx\hx\mbox{for}\!\!\hx\hx k<j.$$ 
All $\;k< j\;$ components of  
$\Pi_{2}^{(3j)},\;\Pi_{2}^{(3j+1)}$ and $\Pi_{2}^{(3j+2)}$ 
are proportional to ${(Q_j)}_k$.\\ We get zero overlap between a polynomial 
$\Pi_Q^{(L)}$ of degree $b_{L,Q}$ with all polynomials $\Pi_{Q'}^{(L')}$ 
which have at least $b_{L,Q}$ vanishing low-$k$ components (these can only 
be polynomials which have $b_{L',Q'}>2\,b_{L,Q}$). One of many such relations 
is e.g.
$$\langle\,\Pi_Q^{(3j)}\;|\:\Pi_{Q'}^{(3j\,')}\,\rangle=0\hx\mbox{for}\hx
 Q\,=\,Q'=\,1 \hx\hx \mbox{and}\hx 2j \le j'-1.$$ 
This property may be called "partial orthogonality".\\[2mm]
Further rules, this time valid for all $k$, regard the vanishing of
many components for particular linear combinations:$\;\;$ Defining 
\bea \cQ_+^{(j)}&\equiv&\Pi_1^{(3j)}=9\,c\,R_{j-1}+3\,Q_{j-1}\;
 \! =[\,\alp^{(+)}_0,\;\alp^{(+)}_1,\;\ldots,\;\alp^{(+)}_{2j-1}\,];
 \ny\\ \cQ_-^{(j)}&\equiv&\hx Q_j\,+9\,c\,R_{j-1}-Q_{j-1}\hx 
 =[\,\alp^{(-)}_0,\;
 \alp^{(-)}_1,\;\ldots,\;\alp^{(-)}_{2j}\:], \ny\eea
we have checked up to $j=30$ that $\;\alp^{(\pm)}_k=0$ for $k<j$ and that 
all even (odd) $k$ Jacobi components of $\cQ_+^{(j)}$ ($\cQ_-^{(j)}$) vanish. 
So we conjecture for all $\;j,\;j\,'\;$   
$$\langle\,\cQ_+^{(j)}\,|\,\cQ_-^{(j\,')}\,\rangle=0. $$
One can check some special cases of these results in Table 1.
It has been found numerically \cite{GeRo} that a similar partial 
orthogonality appears also for the $\Zmb_N$-Baxter polynomials with 
$N=4,\:5,\:6.$ For even values of $N$ 
further relations emerge, but these will not be discussed here.
   
\section{Conclusion}
The polynomials which play a central role for the calculation of the
energy eigenvalues of the superintegrable $\Zmb_3-$chiral Potts model are 
found to be related to Jacobi polynomials in a very peculiar way. Many 
integrals giving the Jacobi-coefficients of Baxter's polynomials are found
to vanish. These observations should have a deeper group-theoretical 
background, but the underlying symmetry is not yet clear to us. 
By reducing the formulation of the problem to some basic facts, the present 
analysis tries to prepare the ground for clarifying the symmetry involved. 

\subsection*{Acknowledgements}
The author is grateful to \,Prof. Mo-Lin\, Ge \,for the very friendly 
hospitality during\\[1mm] the APCTP 2001 Nankai Symposium in Tianjin. 
This work has been supported by \\[1mm] INTAS-OPEN-00-00055.

\end{document}